\documentclass[pra,aps,floatfix,showkeys,tightenlines,superscriptaddress,amsmath,amssymb,10pt,nofootinbib]{revtex4-2}
\usepackage{slashed}
\usepackage{xcolor}
\usepackage{graphicx}
\usepackage{hyperref}

\newcommand{\be}{\begin{equation}}\newcommand{\ee}{\end{equation}}
\newcommand{\bea}{\begin{eqnarray}}\newcommand{\eea}{\end{eqnarray}}
\newcommand{\brr}{\begin{array}}\newcommand{\err}{\end{array}}
\newcommand{\bit}{\begin{itemize}}\newcommand{\eit}{\end{itemize}}
\newcommand{\ben}{\begin{enumerate}}\newcommand{\een}{\end{enumerate}}

\newcommand{\bbm}{\begin{bmatrix}}\newcommand{\ebm}{\end{bmatrix}}
\newcommand{\ba}{\begin{array}}
\newcommand{\ea}{\end{array}}
\newcommand{\G}{\textbf}

\newtheorem{mydef}{Definition}
\newtheorem{Lemma}{Lemma}
\newcommand{\bd}{\begin{mydef}} \newcommand{\ed}{\end{mydef}}
\newcommand{\bthe}{\begin{theorem}} \newcommand{\ethe}{\end{theorem}}
\newcommand{\ble}{\begin{Lemma}} \newcommand{\ele}{\end{Lemma}}

\newcommand{\dr}{\mathrm{d}}

\def\ha{\frac{1}{2}}

\def\intx{\int \!\!\mathrm{d}^3 {\G x}}

\def\ph{\varphi}

\def\lab{\label}\def\lan{\langle}
\def\lf{\left}

\def\non{\nonumber}\def\pa{\partial}\def\ran{\rangle}
\def\rar{\rightarrow}
\def\ri{\right}
\def\al{\alpha}\def\bt{\beta}\def\ga{\gamma}\def\Ga{\Gamma}
\def\de{\delta}\def\De{\Delta}

\def\la{\lambda}\def\si{\sigma}\def\Si{\Sigma}
\def\om{\omega}\def\Om{\Omega}

\def\mass{{_{1,2}}}
\def\flav{{e,\mu}}\def\1{{_{1}}}\def\2{{_{2}}}

\def\noHe0{:\;\!\!\;\!\!:H_e(0):\;\!\!\;\!\!:}
\def\noHm0{:\;\!\!\;\!\!:H_\mu(0):\;\!\!\;\!\!:}

\def\lab{\label}
\def\lan{\langle}
\def\lf{\left}

\def\non{\nonumber}
\def\pa{\partial}\def\ran{\rangle}
\def\rar{\rightarrow}
\def\ri{\right}

\def\al{\alpha}\def\bt{\beta}\def\ga{\gamma}
\def\Ga{\Gamma}\def\de{\delta}\def\De{\Delta}

\def\la{\lambda}
\def\si{\sigma}\def\Si{\Sigma}
\def\om{\omega}\def\Om{\Omega}

\def\mass{{_{1,2}}}
\def\flav{{e,\mu}}\def\1{{_{1}}}\def\2{{_{2}}}

\begin{document}

\title{Time-energy uncertainty relation for neutrino oscillations: historical development, applications and future prospects}

\author{G.~G.~Luciano}
\email{giuseppegaetano.luciano@udl.cat}
\affiliation{Department of Chemistry, Physics and Environmental and Soil Sciences, Escola Polit\`ecnica Superior, Universitat de Lleida, Av. Jaume II, 69, 25001 Lleida, Spain}

\author{L.~Smaldone}
\email{lsmaldone@unisa.it}
\affiliation{Dipartimento di Fisica, Universit\`a degli Studi di Salerno, Via Giovanni Paolo II, 132 I-84084 Fisciano (SA), Italy.}
\affiliation{INFN Gruppo Collegato di Salerno - Sezione di Napoli c/o Dipartimento di Fisica, Universit\`a di Salerno, Italy.}

\date{\today}

\begin{abstract}
Time-energy uncertainty relation (TEUR) plays a fundamental role in quantum mechanics, as it allows to grasp peculiar aspects of a variety of phenomena based on very general principles and symmetries of the theory. 
Using the Mandelstam-Tamm method, TEUR has been recently derived for neutrino oscillations by connecting the uncertainty on neutrino energy with the characteristic time-scale of oscillations. Interestingly enough, the suggestive interpretation of neutrinos as unstable-like particles has proved to naturally emerge in this context. 
Further aspects have been later discussed in semiclassical gravity by computing corrections to the neutrino energy uncertainty in a generic stationary curved spacetime, and in quantum field theory, where the clock observable turns out to be identified with the non-conserved flavor charge operator. 
In the present work, we give an overview on the above achievements. 
In particular, we analyze the implications of TEUR and explore the impact of gravitational and non-relativistic effects on the standard condition for neutrino oscillations. 
Correlations with the quantum-information theoretic analysis of oscillations
and possible experimental consequences are qualitatively discussed.
\end{abstract}


 \maketitle

\section{Introduction}
\label{Intro} 

Uncertainty relations constitute a cornerstone of quantum theory. On one hand, the Heisenberg principle sets a fundamental lower limit on how precisely we can perform simultaneous measurements of position and momentum of any quantum system. A direct consequence is that an exactly specified (classical-like) phase-space trajectory cannot be defined for these systems. Another such pair of incompatible variables are energy and time. 
An informal meaning of the time-energy uncertainty relation (TEUR) is that quantum states that only exist for a relatively short time cannot have a sharply determined energy.

Despite formal similarities, the analogy between TEUR and the Heisenberg principle cannot be pursued all the way because of the lack of a time operator in quantum mechanics (QM). As a matter of fact, the proper formalization and general validity of TEUR have been  controversial issues since the advent of quantum theory. 
For a comprehensive review on the subject, one can refer to~\cite{Bauer:1978wd}, while a recent discussion on the non-uniqueness of TEUR formulation appears in~\cite{Busch:2001akq}. In what follows, we shall basically resort to the most general Mandelstam--Tamm version of TEUR, which is based on the only assumption that the evolution of Heisenberg operators in quantum theory is ruled by the Hamiltonian $H$~\cite{ManTam}. Using natural units $\hbar=c=G=1$, TEUR in this framework takes the form
\be
\label{TEURMT}
\Delta E \, \Delta t \, \geq \frac{1}{2} \,, 
\ee
where $\Delta E\equiv\sigma_H$ is interpreted as the standard deviation of the energy on a generic quantum state $|\psi\rangle$, and 
\be
\Delta t \equiv \si_O/\lf|\frac{\dr \lan O(t) \ran}{\dr t}\ri| \, .
\label{teunc0}
\ee
Here $O(t)$ denotes the ``clock observable'', whose dynamics quantifies temporal changes of the system, and $\Delta t$ is the characteristic time-scale over which the mean value $\langle O(t)\rangle\equiv\langle\psi|O(t)|\psi\rangle=\langle\psi(t)|O|\psi(t)\rangle$ varies by a standard deviation.

One of the most common applications of TEUR concerns the understanding of the decay of excited states of atoms. In this scenario, 
the minimum time it takes for an atom to decay to its ground state and radiate is related to the energy uncertainty of the excited state~\cite{Wigner1988}. Based on the Mandelstam-Tamm method, in~\cite{Bilenky:2005hv} TEUR was derived for neutrino flavor oscillations in the QM picture. 

Neutrino oscillations is a phenomenon characterized by a finite time scale, the oscillation time~\cite{giunti2007fundamentals} and, thus, particularly suited for a description in terms of Eq.~\eqref{TEURMT}. 
By identifying $\Delta t$ with the time during which a significant change of the flavor content happens for a given neutrino state, the model of~\cite{Bilenky:2005hv} came to the conclusion that a non-vanishing neutrino energy uncertainty $\Delta E$ is actually needed for oscillations to occur. 

Interesting advances in the study of TEUR for neutrino oscillations were later achieved from both the phenomenological and theoretical points of view. In the former context, the applicability of TEUR to the M\"ossbauer neutrino experiment was analyzed with conflicting results~\cite{Bilenky2008,Akhmedov:2008zz}. On the theoretical side, a generalized TEUR was obtained in~\cite{Blasone2020} for a generic stationary curved spacetime, while in~\cite{Blasone2019} Eq.~\eqref{TEURMT} was recast in terms of a flavor-energy uncertainty relation by using the quantum field theoretical (QFT) formalism of~\cite{Blasone:1995zc,Blasone:2001du}. Remarkably, the 
QFT approach of~\cite{Blasone2019} brings out the interpretation of TEUR for flavor neutrinos as fundamental bounds on energy-variances, in close analogy with the case of
unstable particles. All of the above considerations make it clear that a careful study of TEUR might reveal valuable information about the true nature of neutrinos and 
the related mechanism of flavor oscillations, which are still subject of active investigation.

Based on the outlined premises, here we review recent progress in the description of TEUR for neutrino oscillations. After a brief historical overview, we summarize basic concepts and results in literature, with focus on the future prospects and research lines that can be inspired by the present analysis. In detail, the structure of the work is as follows: in Section \ref{Bilenky}, we go through the derivation of the original QM TEUR for both neutrinos and unstable particles. Section~\ref{Gravity} is devoted to explore gravitational effects on TEUR, while the extension to QFT is investigated in Section~\ref{QFT}. Conclusions and discussion are finally summed up in Section~\ref{Disc}. 

\section{TEUR for oscillating and unstable particles}
\label{Bilenky}

Let us start by presenting TEUR for neutrino oscillations as first derived by Bilenky~\cite{Bilenky:2005hv}. The result is compared to the corresponding formula for unstable particles~\cite{BhattaQD,Blasone:2020qbo}, emphasizing conceptual similarities and differences. 

\subsection{TEUR for neutrino oscillations (\`a-la Bilenky)} \label{bilteur}
Following~\cite{Bilenky:2005hv}, we set $O=P_l=|\nu_l\rangle\langle\nu_l|$ in the Mandelstam-Tamm inequality~\eqref{TEURMT}, 
where $|\nu_l\rangle$ ($l=e,\mu,\tau$) is the flavor neutrino state ($\langle \nu_l|\nu_{l'}\rangle=\delta_{ll'}$) and $P_l$ the projection operator on this state. 
It is straightforward to check that the average value of $P_l$ on a generic state $|\psi(t)\rangle$ is nothing but the 
probability to find the flavor neutrino $\nu_l$ in $|\psi(t)\rangle$, i.e. $\langle P_l(t)\rangle=|\langle\nu_l|\psi(t)\rangle|^2$. Hence, assuming $|\psi(0)\rangle=|\nu_l\rangle$, one has that $\langle P_l(t)\rangle$ coincides with the survival probability $P_{\nu_l\rightarrow\nu_l}(t)$ of the neutrino $\nu_l$ at time $t$. Clearly, 
the following relations hold true: 
\be
P_{\nu_l\rightarrow\nu_l}(0)=1\,,\quad\,\,P_{\nu_l\rightarrow\nu_l}(t)\le1\,\,\,\mathrm{for\,\,any}\,\, t>0\,.
\ee

Projection operators are idempotent, which means they satisfy $P^2=P$. From this condition, it follows that the standard deviation $\Delta P_l$ obeys
\be
\Delta P_l(t)=\sqrt{P_{\nu_l\rightarrow\nu_l}(t)-P^2_{\nu_l\rightarrow\nu_l}(t)}\,.
\ee
By substitution into Eq.~\eqref{TEURMT}, we derive
\be
\label{BilTEUR}
\Delta E\ge\frac{1}{2}\hspace{0.2mm}\frac{|\frac{d}{dt}P_{\nu_l\rightarrow\nu_l}(t)|}
{\sqrt{P_{\nu_l\rightarrow\nu_l}(t)-P^2_{\nu_l\rightarrow\nu_l}(t)}}\,.
\ee

Let us now consider the survival probability $P_{\nu_l\rightarrow\nu_l}(t)$ in the time interval $0 \leq t \leq t_{1min}$,
where $t_{1min}$ is the time when $P_{\nu_l\rightarrow\nu_l}(t)$ reaches the first minimum. In this range,  $P_{\nu_l\rightarrow\nu_l}(t)$ is a monotonically decreasing function~\cite{Bilenky:2005hv,Bilenky2008,Akhmedov:2008zz,Bilenky:2008dk,Bilenky:2009zz,Bilenky:2011pk}. By integration of Eq.~\eqref{BilTEUR} from $0$ to $t$, we finally get
\be 
\label{final}
\Delta E\, t\ge \frac{1}{2}\left[\frac{\pi}{2}-\sin^{-1}\left(2P_{\nu_l\rightarrow\nu_l}(t)-1\right)\right].
\ee

It is useful to apply the above inequality to some specific cases of experimental interest. For instance, atmospheric neutrinos are produced by cosmic rays interacting in the upper atmosphere of the Earth. In turn, the resulting pions and kaons decay into muon neutrinos and muons, many of which give rise to electrons and a muon neutrino--electron neutrino pair. A crude estimation indicates that muon neutrinos are roughly two times more abundant than electron neutrinos~\cite{K2K:2004iot,Super-Kamiokande:2004orf}. 
In the atmospheric Long Baseline region, the survival probability $P_{\nu_\mu\rightarrow\nu_\mu}(t)$ is driven by the neutrino mass-squared difference $\Delta m_{23}^2$ (for more details on the theoretical treatment of neutrino mixing, see Sec.~\ref{Gravity} and Sec.~\ref{QFT}). 
By taking into account that $P_{\nu_\mu\rightarrow\nu_\mu}(t^{(23)}_{1min})\simeq0$, where $t^{(23)}_{1min}=2\pi E/\Delta m_{23}^2$,
we obtain the following TEUR for ultrarelativistic neutrinos ($L\simeq t$)~\cite{Bilenky:2005hv}
\be
\label{atm}
\Delta E\, t_{osc}^{(23)}\ge\pi\,,
\ee
where $t_{osc}^{(23)}=2t^{(23)}_{1min}$ denotes the period of oscillations in the atmospheric-Long Baseline region. As explained in~\cite{Bilenky:2005hv}, the above relation provides a necessary condition for atmospheric neutrino oscillations to be observed.

As a further example, we consider the $\Delta m_{23}^2$-driven survival probability of $\bar\nu_e$. According to CHOOZ experiment~\cite{CHOOZ:1999hei}, this probability is close to the maximal value of unity. Using the condition $P_{\bar\nu_e\rightarrow\bar\nu_e}(t_{1min}^{(23)})=1-\sin^22\theta_{13}$ along with the constraint $\sin^22\theta_{13}\lesssim2\cdot 10^{-1}$, where $\theta_{13}$ is the mixing angle, 
Eq.~\eqref{final} becomes (up to terms of order $\sin^22\theta_{13}$)
\be
\Delta E\, t_{osc}^{(23)}\ge2\sin2\theta_{13}\,,
\ee
which is less stringent than the bound~\eqref{atm}.

{An interesting experimental application of TEUR in high-energy regime concerns Mossbauer neutrino effects~\cite{Kells:1983iac,Bilenky:2008dk, Bilenky2008,Akhmedov:2008zz,Akhmedov:2008jn,Bilenky:2009zz,Machado:2011tn,Raghavan:2012sy}. Mossbauer neutrinos could be produced and detected in the recoiless Mossbauer transitions
\be
^{3}H\rightarrow\hspace{0.3mm} ^{3}\hspace{-0.3mm}He +\bar\nu_e\,,\quad  ^{3}He +\bar\nu_e\rightarrow\hspace{0.3mm} ^{3}\hspace{-0.3mm}H\,,
\ee
with energy $E_\nu\simeq18.6\,\mathrm{keV}$ and relative uncertainty $\Delta E_{\nu}/E_{\nu}\simeq 4.5\times10^{-16}$. 
Flavor oscillations of Mossbauer neutrinos 
were studied in~\cite{Akhmedov:2008jn} by 
using the macroscopic Feynman-diagrams approach~\cite{Giunti:1993se,Beuthe:2001rc,Akhmedov:2008jn,Naumov:2009zza,Naumov:2010um}. This result was, however, challenged by Bilenky \emph{et al.}~\cite{Bilenky2008}, who estimated that the minimum uncertainty needed for Mossbauer oscillations to occur according to TEUR should be $\Delta E_{\nu}/E_{\nu}\gtrsim 10^{-12}$ for atmospheric neutrinos, $\Delta E_{\nu}/E_{\nu}\gtrsim 10^{-13}$ for reactor neutrinos and $\Delta E_{\nu}/E_{\nu}\gtrsim 10^{-14}$ for solar neutrinos. This controversy was later debated in~\cite{Bilenky:2008dk,Akhmedov:2008zz,Akhmedov:2008jn,Bilenky:2009zz}. In particular, a connection between this issue and the possible non-stationary nature of oscillations was investigated in~\cite{Bilenky:2008dk,Bilenky:2009zz}. The conclusion was that, if time plays an active role in neutrino oscillations, one cannot escape from implications of TEUR. On the other hand, if oscillations are a stationary phenomenon, the viewpoint of Akhmedov~\cite{Akhmedov:2008jn} would be the correct one. In the absence of a fundamental and widely shared theory of neutrino
oscillations, valuable hints could be provided by the possible future detection of Mossbauer neutrinos.}

\subsection{TEUR and unstable particles}
\label{Unstable}

We here digress briefly to discuss TEUR for unstable particles. Let us stress that such systems provide one of the most interesting examples where TEUR is typically applied, see e.g.~\cite{griffiths2017introduction}. Toward this end, we stick to the original treatment of~\cite{BhattaQD} and the recent review in~\cite{Blasone:2020qbo}. 

The analysis of~\cite{BhattaQD}, which inspired the derivation of TEUR for neutrino oscillations, is built upon the same considerations leading to Eq.~\eqref{BilTEUR}, with $P_{\nu_l\rightarrow\nu_l}$ being now replaced by the so-called quantum non-decay probability $P_t$. For convenience, let us recast the inequality~\eqref{BilTEUR} as 
\be
\label{BatTEUR}
\left|\frac{d}{dt}P_t\right|\le 2 \Delta E \sqrt{P_t-P_t^2}\,,
\ee
which allows to identify the particle half-life $T_h$, for which $P_t=1/2$ and the right-hand side attains its maximum value. Then, Eq.~\eqref{BatTEUR} gives the weaker inequality
\be
\label{BatTEURbis}
\left|\frac{d}{dt}P_t\right|\le \Delta E\,,
\ee
which sets a time-limit to the instability of decaying quantum systems (see below for more quantitative discussion).

Since for decaying particles $P_{t=0}=1$ and $P_{t\rightarrow\infty}\rightarrow0$, we can infer the following features that are likely to be of interest in the asymptotic regimes of decay
\begin{eqnarray}
&&\left|\frac{d}{dt}P_t\right|=0\,,\quad\,\, \mathrm{for}\,\,\,t=0\,,\\[2mm]
&& \frac{d}{dt} \cos^{-1} P_t\le 2 \Delta E\,, \quad\,\, \mathrm{for}\,\,\,t\rightarrow0\,,\\[2mm]
&&\frac{d}{dt} P_t^{1/2}\le \Delta E\,, \quad\,\, \mathrm{for}\,\,\,t\rightarrow\infty\,.
\end{eqnarray}
These can be unified as
\be
\frac{d}{dt} \cos^{-1} P_t^{1/2}\le \Delta E\,, \quad\,\, \mathrm{for}\,\,\,0\le t\le\infty\,.
\ee
Integration of the the latter inequality gives
\be
\label{Batanal}
\Delta E\, t\ge \cos^{-1} P_t^{1/2}\,,
\ee
which is, in fact, the analogue of Eq.~\eqref{final} for unstable particles. 
At this stage, it becomes clear that the parallelism between neutrinos and unstable particles has a merely formal meaning: indeed, although a TEUR can be associated to both systems, the phenomenon of oscillations is not strictly equivalent to a decay, since the oscillation probability does not vanish at asymptotically long time. 
Notice also that Eq.~\eqref{Batanal} fixes the minimum-time limit to $\pi/(2\Delta E)$. 

Now, for $t$ equal to the half-life $T_h$ defined earlier, Eq.~\eqref{Batanal} leads straightforwardly to
\be
\label{Th}
\Delta E\, T_h\ge \frac{\pi}{4}\,\,\quad\,\,\mathrm{for}\,\,\, \,0<T_h\le\frac{\pi}{2\Delta E}\,. 
\ee
On the other side, if $T_h>\pi/(2\Delta E)$, the non-negativity of $P_t$ allows us to write 
\be
P_t\ge0>1-\frac{2\Delta E\,t}{\pi}\,,\quad\,\, \mathrm{for}\,\,\,t>\frac{\pi}{2\Delta E} \, , 
\ee
which again implies the inequality~\eqref{Th}. Combining these two results together, we finally obtain the Mandelstam-Tamm version of TEUR 
for a decaying quantum system
\be
\label{Batlimit}
\Delta E\, T_h\ge \frac{\pi}{4}\,.
\ee

The above derivation has been recently revised in~\cite{Blasone:2020qbo}, leading to a more stringent bound. Based on the approach of~\cite{DeFilippo:1977bk}, the basic idea is to regard a system of unstable particles as an open quantum system. This entails that a comprehensive analysis cannot be limited to the system alone, but must necessarily involve its surrounding ``environment'' too. In a field theoretical language, such a  prescription amounts to saying that it is not sufficient to describe the system as excitations of some fields on a single vacuum state, but one should take into account extra (artificial) degrees of freedom. A typical strategy consists in doubling each physical degree of freedom, resulting into an enlargement of the Hilbert space. Although this construction was originally proposed to develop a QFT at finite temperature -  Thermo Field Dynamics~\cite{umezawa1982thermo,umezawa1993advanced,Blasone:2018ynq}, various applications have been later considered in QFT on curved spacetime~\cite{Israel:1976ur}
and quantum brain model~\cite{Vitiello:1995wv}, among others. 

For practical purposes, let us consider a set of canonical fermionic operators $a_{\G k}$, $a^{\dag}_{\G k}$, which annihilate and create a decaying particle of momentum ${\G k}$ at $t = 0$, respectively.  The associated vacuum is denoted by $|0\rangle$, i.e. $a_{\G k}|0\rangle=0$. Following the standard notation~\cite{umezawa1982thermo,umezawa1993advanced}, we indicate the corresponding set of
\emph{doubled} operators by $\tilde{a}_{\G k}$, $\tilde{a}^{\dag}_{\G k}$ and the related vacuum by $|\tilde{0}\rangle$, so that  $\tilde{a}_{\G k}|\tilde{0}\rangle=0$.
Notice that such new operators obey the same (anti-) commutation relations as $a_{\G k}$, $a^{\dag}_{\G k}$, i.e. $\left\{a_{\textbf{k}},a^\dagger_{\textbf{p}}\right\}=\left\{\tilde a_{\textbf{k}},\tilde a^\dagger_{\textbf{p}}\right\}=\delta\left(\textbf{k}-\textbf{p}\right)$\,.

Using the above tools, it is natural to define an enlarged Fock space having the tensor-product structure between the Fock spaces associated to $a_{\G k}$ and $\tilde a_{\G k}$, respectively. In this setting, the underlying vacuum state $|0\rangle\!\rangle = |0\rangle \otimes |\tilde{0}\rangle$ is defined in such a way that 
\be
a_{\G k}|0\rangle\!\rangle \ = \ \tilde{a}_{\G k}|0\rangle\!\rangle \ = \ 0  \, .
\ee

The core idea of~\cite{Blasone:2020qbo} is to construct a state $|0(\ph)\rangle$,  such that the expectation value of the number operator $N_\G k \ = \ a^{\dag}_{\G k} a_{\G k}$ satisfies
\be \label{numex}
\mathcal{N}_{\G k}(t) \ \equiv \ \langle 0(\ph)| N_\G k|0(\ph)\rangle \ = \ \exp\lf(-\Ga_\G k \, t\ri) \, , 
\ee
where $\Ga_\G k$ is the inverse lifetime of the unstable particle (the physical meaning of the parameter $\varphi$ will become clear later).  Inspired by Thermo Field Dynamics~\cite{umezawa1982thermo}, one can introduce the new set of ladder operators via the Bogoliubov transformations
\bea \label{bog1}
a_\G k(\ph) & = & \, \cos \ph_{\G k} \, a_\G k \, - \, \sin \ph_{\G k} \, \tilde{a}^\dag_\G k \, , \\[2mm] \label{bog2}
\tilde{a}_\G k(\ph) & = & \, \cos \ph_{\G k} \, \tilde{a}_\G k \, + \, \sin \ph_{\G k} \, a^\dag_\G k \, .
\eea
The related vacuum is defined by the usual condition $a_\G k(\ph) \, |0(\ph)\rangle = \tilde{a}_\G k(\ph) \, |0(\ph)\rangle=0$. By explicit computation, it is possible to show that~\cite{umezawa1982thermo}
\be \label{unvac}
|0(\ph)\rangle \ = \ \prod_{\G k} \lf(\cos \ph_{\G k} \, + \, \sin \ph_{\G k} a^\dag_\G k \, \tilde{a}^\dag_\G k \ri) \, |0\rangle \!\rangle \, .
\ee
Combining Eqs.~\eqref{numex} and~\eqref{unvac}, we infer 
\be \label{expdec}
\mathcal{N}_{\G k}(t) \ = \ \sin^2 \ph_{\G k}(t) \ = \ \exp\lf(-\Ga_{\G k}t\ri) \, ,
\ee
which displays the meaning of $\ph_{\G k} \equiv \ph_{\G k}(t)$ in terms of the inverse lifetime $\Gamma_{\G k}$.  

To have more physical intuition on $|0(\ph)\rangle$, let us consider the asymptotic regions $t\rightarrow0$ and $t\rightarrow\infty$. In the first case, we have 
\be \label{vac0}
\lf. |0(\ph)\rangle \ri|_{t=0} \ = \ \prod_{\G k} \, |0(\ph_\G k)\rangle \, , 
\ee
where
\be
|0(\ph_\G k)\ran \ = \  a^\dag_\G k \, \tilde{a}^\dag_\G k \, |0\ran\!\ran \, ,
\ee
is the state describing an unstable particle of momentum $\textbf{k}$.
On the other hand, the limit
\be
\lf. |0(\ph)\ran \ri|_{t=\infty} \ = \ |0\ran\!\ran \, ,
\ee
gives the zero-particle state.

In order to extract TEUR, we now switch to the Heisenberg representation. In this picture, 
the expectation value~\eqref{expdec} is kept unchanged, while the vacuum $|0(\ph)\ran$ must be set as in Eq.~\eqref{vac0}.
In turn, the number operator $N_\G k$ defined above Eq.~\eqref{numex} gains a non-trivial time-dependence. With this setting, we can choose $O(t) = N_{\G k}(t)$ in Eqs.~\eqref{TEURMT} and~\eqref{teunc0} to obtain
\be
\si^2_N \ =  \ \langle 0(\ph)| N^2_\G k(t)|0(\ph) \rangle \ - \ \langle 0(\ph)| N_\G k(t) |0(\ph)\rangle^2 \ = \ \mathcal{N}_{\G k}(t)\lf(1-\mathcal{N}_{\G k}(t)\ri) \, ,
\ee
which yields 
\be \label{neutunqm1}
\lf|\frac{\dr\mathcal{N}_{\G k}(t)}{\dr t}\ri| \,\leq  \,2\Delta E  \,\sqrt{\mathcal{N}_{\G k}(t)\lf(1-\mathcal{N}_{\G k}(t)\ri)} \, .
\ee
Notice that this inequality is the same as Eq.~\eqref{BatTEUR}, with the non-decay probability $P_t$ being now formally replaced by the vacuum density $\mathcal{N}_{\G k}(t)$. 
For the characteristic time $T_h$ that maximizes the right side, we have $\mathcal{N}_{\G k}(T_h)=1/2$ and 
\be \label{wineq}
\Delta E \, \geq \, \lf|\frac{\dr\mathcal{N}_{\G k}(t)}{\dr t}\ri| \, ,
\ee
to be compared with Eq.~\eqref{BatTEURbis}.

The relation~\eqref{wineq} can be further manipulated by resorting to the triangular inequality and integrating both
sides from $0$ to $T$ to obtain
\be
\label{31}
\Delta E \, T  \ \geq  \ \int^T_{0} \!\! \dr t \, \lf|\frac{\dr\mathcal{N}_{\G k}(t)}{\dr t} \ri| \, \geq  \,\lf|\int^T_{0} \!\! \dr t \, \frac{\dr\mathcal{N}_{\G k}(t)}{\dr t} \ri| \, ,
\ee
which yields
\be \label{etrel1}
\Delta E\,  T \geq 1-\mathcal{N}_{\G k}(T)   \, .
\ee
At $T=T_h$, we then get the Heisenberg-like lower bound
\be \label{teurHeis}
\Delta E\, T_h \geq \ha \, ,
\ee
that is stronger than the one in Eq.~\eqref{Batlimit}. 

As a final remark, it is worth noting from Eq.~\eqref{expdec} that $T_h= \log 2/{\Ga_{\G k}} = \tau_{\G k} \, \log 2$, where $\tau_{\G k}=1/{\Ga_{\G k}}$ is the particle life-time. Therefore, TEUR~\eqref{teurHeis} can be rearranged as
\be \label{teurun}
\Delta E \geq \frac{1}{\tau_{\G k} \, \log 4} \, .
\ee
This inequality sets an intrinsic lower bound on the energy distribution width of unstable quantum particles. In the same way, by carrying on the analogy between neutrinos and unstable particles, we can interpret Eq.~\eqref{BilTEUR} as a fundamental limit on the width of the neutrino energy distribution.

\section{Gravitational effects on TEUR}
\label{Gravity}

In~\cite{Blasone2020} TEUR for neutrino oscillations has been derived for a generic stationary curved spacetime to study how gravity affects the characteristic oscillation length. 
To review such a formalism, we first take a step back and consider the Bilenky-like (Minkowski) form~\eqref{final} of TEUR, here rewritten equivalently as
\be
\label{BilTEURNew}
2\Delta E{\sqrt{P_{\nu_l\rightarrow\nu_l}(x)-P^2_{\nu_l\rightarrow\nu_l}(x)}}\ge\left|\frac{d}{dt}P_{\nu_l\rightarrow\nu_l}(x)\right|
\,,
\ee
where, for more generality, we have assumed that the neutrino survival probability $P_{\nu_l\rightarrow\nu_l}$  
depends on $t$ through the spacetime coordinate $x\equiv x(t)$. For simplicity, from now on we restrict to a toy model involving two flavors only, but the same considerations can be extended to the more realistic case of three neutrino generations.

As observed in Sec.~\ref{Unstable}, the left side of Eq.~\eqref{BilTEURNew} reaches its maximum
when $P_{\nu_l\rightarrow\nu_l}=1/2$. 
Integrating both sides between some initial time $t_0$ (such that $P_{\nu_l\rightarrow\nu_l}(x(t_0))=1$) and $t$, 
and employing once more the triangular inequality, we are led to~\cite{Blasone2020}
\begin{equation}
\label{triang}
\Delta E \, T  
\geq  
P_{\nu_l\rightarrow\nu_{l'}}\hspace{0.5mm}(x(t))\,,
\end{equation}
where $T \equiv \int_{t_0}^{t}dt'=t-t_0$
is the neutrino flight-time and $P_{\nu_l\rightarrow\nu_{l'}}\hspace{0.5mm}(x(t))=1-P_{\nu_l\rightarrow\nu_{l}}\hspace{0.5mm}(x(t))$ the oscillation probability. 

Gravitational corrections to flavor oscillations typically enter the neutrino phase~\cite{ahluwalia,fuller,epl,fornengo,capozziellobrans,Capozziello:2011et,piriz,pulsar,extended,dvornikov,mottola,rev,Lambiase:2005gt,Visinelli:2014xsa,Lambiase:2022ucu,Luciano:2021gdp,CapQuar,CapQuarbis}. To find out the form of these extra terms, we write explicitly the generic flavor eigenstate
$|\nu_l\rangle$ as a mixture of the mass eigenstates $|\nu_k\rangle$ ($k=1,2$) according to~\cite{giunti2007fundamentals}
\be \label{eqn:genstate}
|\nu_l(x)\rangle =  \sum_{k=1,2} U_{{l k}}\hspace{0.1mm}(\theta) \, |\nu_k(x)\rangle \,,\quad\,\, l=e,\mu\,, 
\ee
where $U_{{l k}}\hspace{0.1mm}(\theta)$ 
is the generic element 
of the mixing matrix~\cite{giunti2007fundamentals}. 

Since the states $|\nu_k(x)\rangle$ are Hamiltonian eigenstates, in the standard quantum theory on flat background their evolution is governed by the usual relation
\begin{equation}
\label{defmas}
|\nu_k({x})\rangle=\exp[{-i\varphi_k(x)}]\hspace{0.2mm}|\nu_k\rangle,\quad (k=1,2)\,,
\end{equation}
where
\begin{equation}
\label{phase}
\varphi_k(x)\equiv E_k\hspace{0.4mm}t-\textbf{p}_k\cdot\textbf{x}
\end{equation}
denotes the QM phase of the $k^{th}$ eigenstate 
having mass $m_k$, energy $E_k$ and three-momentum $\textbf{p}_k$, respectively (not to be confused with the parameter $\ph_{\G k}$ appearing in the Bogoliubov transformations~\eqref{bog1} and~\eqref{bog2}). 
These three quantities are connected via the flat mass-shell relation $E^2_k=m^2_k+{|\textbf{p}_k|}^2$. 

For relativistic neutrinos traveling along the $x$-axis from the source point $x_0$ to the detector point $x$ (with $x>x_0$), the phase~\eqref{phase} can be simplified to
\begin{equation}
\label{eqn:phasebis}
\varphi_{k}\simeq\frac{m^2_k}{2E
}\hspace{0.2mm}L_p\,,
\end{equation}
where we have assumed the mass eigenstates to have roughly the same energy $E$. Furthermore, we have denoted the proper distance traveled by neutrinos by $L_p=x-x_0$, which can be approximated to the flight-time $T$ in the relativistic limit. 

Now, substitution of Eq.~\eqref{eqn:phasebis} into~\eqref{eqn:genstate} and~\eqref{defmas} gives
\begin{equation}
\label{relapp}
|\nu_l({x})\rangle=
\sum_{k=1,2}U_{{l k}}\hspace{0.1mm}(\theta)\hspace{0.2mm}\exp\left({-i\hspace{0.3mm}\frac{m^2_k}{2E
}\hspace{0.3mm}L_p}\right)\hspace{-0.4mm}|\nu_k\rangle\,, \quad\,\, l=e,\mu\,.
\end{equation}
In turn, the probability that a neutrino 
$|\nu_l\rangle$ undergoes a flavor transition to
$|\nu_{l'}\rangle$ after propagating over the distance $L_p$ is~\cite{Blasone2020}
\begin{eqnarray}
\nonumber
P_{\nu_l\rightarrow \nu_{l'}}(x)&=& |\langle \nu_{l'}(x)|\nu_l(x_0)\rangle|^2\\[2mm]
\nonumber
&=& \sin^2 (2 \theta)\, \sin^2 \lf(\frac{\varphi_{12}}{2}\ri)\\[2mm]
&=& \sin^2 (2 \theta)\sin^2 \lf(\frac{\pi\hspace{0.2mm}L_p}{L^{osc}}\ri)\,,\quad\,\,\, l\neq l'\,,
\label{ProbOver}
\end{eqnarray}
where in the second step $\varphi_{12}\equiv \varphi_1-\varphi_2$ denotes the phase-shift acquired by the mass eigenstates during the propagation, while in the last step  we have introduced the characteristic \emph{oscillation length} $L^{osc}$ 
\be
\label{oscleng}
L^{osc}\equiv\frac{4 \pi E
}{\De m^2}\,, \quad\,\,\, \Delta m^2=|m_1^2-m_2^2|\,.
\ee

Accordingly, TEUR in Eq.~\eqref{triang} takes the form
\be
\label{teurriformulation}
\Delta E \, T  
\geq  \sin^2 (2 \theta)\sin^2 \lf(\frac{\pi \hspace{0.2mm} L_p}{L^{osc}}\ri).
\ee
Since this inequality holds for any spacetime point, we can set $L_p=L^{osc}/2$ so  
as to maximize the right side and obtain the neutrino oscillation condition
\be \label{minkteur}
\De E \geq \frac{2 \, \sin^2(2 \theta)}{L^{osc}} \, .
\ee

Two comments are in order here: first, we notice that for $\theta=0$ and/or $L^{osc}\rightarrow\infty$, the lower bound on the neutrino energy uncertainty vanishes, as expected in the absence of mixing and/or oscillations. Moreover, in~\cite{Ahluwalia:1996fy}
$\Delta E$ is interpreted as the minimum energy transferred to neutrinos in scattering processes, which is necessary to reveal them once the oscillation occurred.

\subsection{TEUR in curved spacetime}

Let us extend the above formalism to stationary curved spacetime. For this kind of metrics, we remark that TEUR can be consistently defined, since there always exists a global timelike 
Killing vector field $K^\mu$, such 
that $K \equiv  \int K_\mu\, T^{\mu \nu}\, \dr \Si_\nu $ does not depend on the spacelike hypersurface $\Sigma$, where  $T^{\mu\nu}$ is the (conserved) stress-energy
tensor~\cite{DeWitt:1975ys}. We are then allowed to introduce a coordinate $t$, such that the metric is independent of it and with respect to which $K^{\mu}=(1,0,0,0)$. One has $K \equiv  \int \sqrt{-g} :T^0_{\,\,\,0}: \dr^3x=H$, where $:T^0_{\,\,\,0}:$ denotes the normal-ordered operator, $g$ is the determinant of the metric tensor $g^{\mu\nu}$ and $H$ the Hamiltonian of the considered system. More generally, the role of the above Killing vector can be taken by Kodama vector for dynamical metrics~\cite{Kodama:1979vn}.  

To generalize TEUR for flavor oscillations to curved spacetime, it proves convenient to rephrase the evolution~\eqref{relapp} of the neutrino state in a manifestly covariant way as~\cite{fuller}
\be
\label{covfor}
|\nu_l({\lambda})\rangle=\sum_{k=1,2} U_{{l k}}\hspace{0.1mm}(\theta)\hspace{0.2mm}\exp(-i\Phi)|\nu_k\rangle\,,\quad \,\,\, l=e,\mu\,,
\ee
where $\lambda$ is the neutrino world-line parameter and $\Phi$ generalizes the definition~\eqref{phase} of neutrino phase to 
\be
\label{phasep}
\Phi=\int_{\lambda_0}^\lambda P_{\mu}\frac{\dr x^\mu_{null}}{\dr \lambda'}\hspace{0.2mm}\dr\lambda'\,.
\ee
Notice that $P_\mu$ and $\dr x^\mu_{null}/d\lambda$ in the above integral are the generator of spacetime translations of neutrino mass states and the null tangent vector to the neutrino worldline, respectively. For $g_{\mu\nu}=\eta_{\mu\nu}$, one can check that Eq.~\eqref{phasep} reproduces the evolution~\eqref{relapp} in flat spacetime, as expected~\cite{Blasone2020}. 

For a generic stationary curved spacetime, the operator $P_\mu$ is defined by the generalized mass-shell relation~\cite{fuller}
\begin{equation}
\label{eqn:genmascond}
\left(P^{\mu}+\frac{A^\mu\gamma^5}{2}\right)\left(P_{\mu}+\frac{A_{\mu}\gamma^5}{2}\right)=M^2\,,
\end{equation}
where $M=\mathrm{diag}[m_1,m_2]$ is the diagonal neutrino mass-matrix and $\gamma_\mu$ are the Dirac matrices obeying the Clifford
algebra $\{\gamma^{\mu},\gamma^{\nu}\}= 2 g^{\mu\nu}$ ($\gamma^5=i\gamma^0\gamma^1\gamma^2\gamma^3$). Furthermore, the potential-like term
\begin{equation}
\label{eqn:A}
A^{\mu}=\frac{\sqrt{g}}{4}\hspace{0.3mm}\hspace{0.3mm}e^{\mu}_{\hat{a}}\hspace{0.3mm}\epsilon^{{\hat{a}}{\hat{b}}{\hat{c}}{\hat{d}}}\left(\partial_\sigma e_{{\hat{b}}\nu}-\partial_\nu e_{{\hat{b}}\sigma}\right)e^{\nu}_{\hat{c}} e^{\sigma}_{\hat{d}}
\end{equation}
enters the spin-connection in the Dirac equation on curved spacetime~\cite{fuller,Blasone2020}. Here   $e^{\mu}_{\hat{a}}$ are the tetrads connecting general curvilinear and locally inertial coordinates, denoted by greek and (hatted) latin indices, respectively. The symbol $\epsilon^{{\hat{a}}{\hat{b}}{\hat{c}}{\hat{d}}}$ stands for the totally antisymmetric Levi-Civita tensor of component $\epsilon^{{\hat{0}}{\hat{1}}{\hat{2}}{\hat{3}}}=+1$, while $\partial$ indicates ordinary derivative. 

As shown in~\cite{fuller}, the integral in~\eqref{phasep} can be simplified for relativistic neutrinos moving along a null trajectory by neglecting $\mathcal{O}(A^2)$ and $\mathcal{O}{(AM^2)}$-terms. Under these assumptions, we obtain~\cite{Blasone2020}
\be\label{phasep2}
\Phi=\int_{\lambda_0}^\lambda \left(\frac{M^2}{2}-\frac{\dr x^\mu_{null}}{\dr \lambda'}A_{\mu}\gamma^5\,\right)\hspace{0.2mm}\dr\lambda'\,,
\ee
which can be further manipulated by rewriting the line element $d\lambda$ in terms of the differential proper distance $\dr \ell$ at constant time for null trajectories as
\begin{equation}
\label{dl}
\dr\lambda  =  \dr \ell{\left(\hspace{-1mm}-g_{ij}\frac{\dr x^i}{\dr \lambda}\frac{\dr x^j}{\dr \lambda}\right)}^{\hspace{-1mm}-\frac{1}{2}}\hspace{-2mm} =  \dr \ell{\left[g_{00}\lf(\frac{\dr t}{\dr \lambda}\ri)^2\hspace{-1.5mm}+2\hspace{0.3mm}g_{0i}\hspace{0.3mm}\frac{\dr t}{\dr \lambda}\hspace{0.3mm}\frac{\dr x^i}{d\lambda}\right]}^{-\frac{1}{2}}.
\end{equation}

Equation~\eqref{phasep2} provides the crucial ingredient toward reformulating TEUR in curved spacetime. To this end, we recast Eq.~\eqref{triang} in terms of $\lambda$ to obtain
\be
\Delta E \, T(\la)   \geq  
\mathcal{P}_{\nu_l\rightarrow \nu_{l'}}\hspace{0.3mm}(x(\lambda))  \, ,
\ee
where the neutrino flight-time is given by $T(\lambda)\equiv t(\lambda)-t(\lambda_0)$ for a fixed $\lambda_0$. From Eq.~\eqref{ProbOver}, it follows that
\be\label{sp}
\Delta E \geq \frac{\mathrm{sin}^2\lf(2\theta\ri)}{T(\lambda)}\,\mathrm{sin}^2\bigg[\frac{{\varphi_{12}(x(\lambda))}}{2}\bigg], \quad {\varphi}_{12}\equiv{\varphi}_1-{\varphi}_2\,,
\ee
where the neutrino phase $\varphi_k$ ($k=1,2$) is now generally defined as the eigenvalue of $\Phi$ in Eq.~\eqref{phasep2} with respect to the $k^{th}$ mass eigenstate, i.e. $\Phi|\nu_k\rangle=\varphi_k|\nu_k\rangle$. 

Before moving onto the evaluation of the above inequality for some specific metrics, it is worth observing that $\Delta E$ appearing in Eq.~\eqref{sp} represents the uncertainty on the neutrino energy as measured by an asymptotic observer (see~\cite{Blasone2020} for more details). However, the truly measurable quantity in neutrino oscillation experiments is the local energy uncertainty $\Delta E_\ell$, which represents the uncertainty for a local observer temporarily at rest
in the curved background (and, thus, with respect to the oscillation experiment). The connection between these two quantities is established via the vierbein fields 
\be
\label{locen}
E_\ell(x)\equiv P_{\hat{0}}(x)=e^{\nu}_{\hat{0}}(x)\, P_{\nu}\,.
\ee
Clearly, this transformation must also be implemented into the expression of the phase-shift $\varphi_{12}$, along with the usage of the proper distance traveled by neutrinos. This is an essential step to  
make a consistent comparison with the TEUR~\eqref{minkteur} in the limit when $g_{\mu\nu}\rightarrow\eta_{\mu\nu}$. 

\subsubsection{Schwarzschild spacetime}
\label{Ss}
The Schwarzschild metric in isotropic coordinates $(t,x,y,z)$ and (linearized) weak-field approximation is
\be\label{schw}
\dr s^2=\lf(1+2\phi\ri)\dr t^2-(1-2\phi)\,(\dr x^2+\dr y^2+\dr z^2)\, ,
\ee
where $\phi(r)=-M/r\equiv-M/\sqrt{x^2+y^2+z^2}$
is the gravitational potential and $M$ the spherically symmetric source mass. The only non-trivial components of the tetrads are $e^0_{\hat{0}}=1-\phi$ and $e^i_{\hat{j}}=(1+\phi)\,\de^i_j$, where $\de^i_j$ is the Kronecker delta. 

Because of the particular symmetry of the metric, the potential $A_{\mu}$ in Eq.~\eqref{eqn:A} vanishes. From Eqs.~\eqref{phasep2} and~\eqref{dl},  the resulting neutrino phase-shift $\varphi_{12}$ takes the simple form~\cite{Blasone2020}
\be\label{osas}
\varphi_{12}=\frac{\Delta m^2}{2}\int^\ell_{\ell_0}\frac{1+\phi}{E}\,\dr \ell'\,,
\ee
with $E\equiv P_0=g_{00}\,\dr t/\dr\lambda$.

If we set the frame axes so that the neutrino propagates radially (e.g. along the $x$-direction), we have $\dr \ell=
(1-\phi)\,\dr x$. Then, by taking into account that $E$ is constant along a null trajectory, the integral~\eqref{osas} can be solved straightforwardly to give the following form of TEUR
\be\label{sps}
\Delta E\geq \frac{\mathrm{sin}^2\lf(2\theta\ri)}{T}\,\mathrm{sin}^2\lf[\frac{\Delta m^2\lf(x-x_0\ri)}{4E}\ri].
\ee
In terms of the proper distance \be
L_p\equiv\int^x_{x_0}\hspace{-1mm}\sqrt{-g_{11}}\,\dr x'
=x-x_0+M\ln\lf(\frac{x}{x_0}\ri) \, , 
\ee 
and the local energy~\eqref{locen}, we obtain
\be
\label{noprl}
\frac{\Delta E_\ell}{1-\phi}\geq \frac{\mathrm{sin}^2\lf(2\theta\ri)}{T}\,\mathrm{sin}^2\lf(\frac{\pi L_p}{L^{osc}_{\phi}}\ri),
\ee
where we have introduced the Schwarzschild-corrected proper oscillation length (at the leading order in $\phi$)~\cite{fornengo}
\be
\label{lsmald}
L^{osc}_{\phi}\equiv \frac{4\pi E_\ell}{\Delta m^2}\left[1+\phi+\frac{M}{x-x_0}\ln\left(\frac{x}{x_0}\right)\right]. 
\ee
Here, the gravitational potential is to be regarded as $\phi(x)=-M/x$.

As emphasized below Eq.~\eqref{teurriformulation}, the right side of the inequality~\eqref{lsmald} is maximized for $L_p=L^{osc}_\phi/2$, which in turn gives the following relation between the time interval $T$ and the oscillation length $L^{osc}_{\phi}$~\cite{Blasone2020}
\be
T 
=  \frac{L^{osc}_\phi}{2}\lf[1+\frac{M}{x-x_0}\mathrm{ln}\lf(\frac{x}{x_0}\ri)\ri].
\ee
By plugging into Eq.~\eqref{noprl}, we can finally cast the TEUR in Schwarzschild spacetime as
\be
\label{bsc}
\Delta E_\ell
\geq  \frac{2\,\mathrm{sin}^2(2\theta)}{L^{{osc}}_{eff}(M)}\,,
\ee
where
\be \label{leffs}
L^{{osc}}_{eff}(M) \equiv L^{osc}_\phi\lf[1+\phi+\frac{M}{x-x_0}\mathrm{ln}\lf(\frac{x}{x_0}\ri)\ri]
\ee
is the \emph{effective oscillation length}.

The last form of TEUR~\eqref{bsc} allows for a direct comparison with the corresponding flat expression~\eqref{minkteur}. 
Remarkably, we can see that gravitational corrections increase
$L^{osc}_{eff}$ with respect to the standard oscillation length~\eqref{oscleng}, resulting into a more stringent bound on $\Delta E_{\ell}$. For consistency, one can check that $L^{{osc}}_{eff}(M)\rightarrow L^{osc}$ for $M\rightarrow0$ and the inequality~\eqref{minkteur} is recovered. On the other hand, in the limit $L^{{osc}}_{eff}(M)\to \infty$, the right side of Eq.~\eqref{bsc} vanishes, in line with the discussion below Eq.~\eqref{minkteur}.

\subsubsection{Lense-Thirring spacetime}

We can now apply the same line of reasoning to extract TEUR in Lense-Thirring background, 
which can be interpreted as the exterior gravitational field of a rotating spherical source. Notice also that this metric is usually adopted to describe
gravitomagnetic frame-dragging effects, see for instance~\cite{ciufolini}.

In the weak-field approximation, the line element on the equatorial plane reads~\cite{Thirring2014OnTI}
\begin{equation}
\label{LeTh}
\dr s^2=\lf(1+2\phi\ri)\dr t^2-\frac{\phi\hspace{0.2mm}\Omega}{r^2}\lf(x \, \dr y-y\, \dr x\ri) \, \dr t-(1-2\phi)\,(\dr x^2+\dr y^2+\dr z^2),
\end{equation}
where $\Omega\equiv4R^2\omega/5$, 
$\omega$ is the angular velocity of the source of radius $R$ and $\phi$ the gravitational potential defined below Eq.~\eqref{schw}. Without loss of generality, we assume the rotation takes place around the $z$ axis. 

Following the same calculations as in Sec.~\ref{Ss}, we now arrive to~\cite{Blasone2020} 
\be \label{teurls}
\De E_\ell \geq \frac{2 \sin^2 (2 \theta)}{L^{osc}_{eff}(M,\Omega)} \, ,
\ee
where the effective oscillation length exhibits a non-trivial dependence on both  the mass and angular velocity of the source as
\begin{equation}
L^{osc}_{eff}(M,\Omega) \equiv  L^{osc}\lf[1+\frac{M}{x-x_0}\mathrm{ln}\lf(\frac{r+x}{r_0+x_0}\ri)+\frac{M\,\Omega}{2b\lf(x-x_0\ri)}\lf(\frac{x}{r}-\frac{x_0}{r_0}\ri)+\phi\left(1+\frac{\Omega b}{r^2}\right)\ri] \, ,
\end{equation}
with $b>R$ being the impact parameter. 
As discussed for Schwarzschild geometry, the effective oscillation length is increased with respect to the flat case. The latter is however restored for $M,\Omega\rightarrow0$, as it should be.

\subsubsection{Rindler spacetime}
As last example, we discuss TEUR for an observer undergoing constant proper acceleration $a>0$ (Rindler observer). It is interesting to note that, although the corresponding Rindler metric is flat, it  well-describes the static gravitational field that would be appropriate for a spacetime with fictitious ``infinite flat earth''~\cite{sciama1981quantum}. 
Assuming the acceleration along the $x$-axis, the (linearized) line element in Rindler-Fermi coordinates is~\cite{Misner:1973prb} 
\be\label{acm}
\dr s^2\ =\ \lf(1+ax\ri)\dr t^2-(\dr x^2+\dr y^2+\dr z^2) \, .
\ee

TEUR for Rindler observer can be cast in the same form as Eq.~\eqref{bsc} (or Eq.~\eqref{teurls}), with the acceleration-dependent oscillation length being now defined by~\cite{Blasone2020}
\be\label{losca}
L^{osc}_{eff}(a) \equiv L^{osc}\lf[1+\frac{a}{4}\lf(x-x_0\ri)\ri].
\ee
Once again, the effective oscillation length is increased with respect to the 
flat case, giving rise to a more stringent bound on the energy uncertainty $\Delta E_\ell$. As expected, the standard formula in Minkowski spacetime is reproduced for $a\rightarrow0$. On other other hand, the limit of $L^{osc}_{eff}(a)\rightarrow\infty$ for fixed $a$ cannot be considered in the present case, as it invalidates the linearized approximation under which Eq.~\eqref{losca} is derived~\cite{Misner:1973prb}.

\section{TEUR in Quantum Field Theory}
\label{QFT}

We start this section by reviewing the TEUR in the Pontecorvo formalism for neutrino mixing and oscillations \cite{Gribov:1968kq,Bilenky:1975tb,Bilenky:1976yj,Bilenky:1977ne,Bilenky:1987ty} as developed in~\cite{Blasone2019}. This analysis gives similar results to the ones shown in Sec.~\ref{Bilenky} and~\ref{Gravity}.
Although the Pontecorvo approach is experimentally successful, it has however several conceptual drawbacks and is widely recognized to be only applicable to relativistic neutrinos~\cite{PhysRevD.37.1935,PhysRevD.45.2414,Blasone:1995zc,Ho:2012yja,Lobanov:2015esa,Fantini:2018itu}. These issues motivate us to introduce the main aspects of the so-called \emph{flavor Fock space} approach to neutrino mixing in QFT~\cite{ALFINITO199591,Blasone:1995zc,PhysRevD.59.113003, Hannabuss:2000hy, PhysRevD.64.013011, PhysRevD.65.096015, Hannabuss:2002cv,Lee:2017cqf,Luciano:2022foe}, which extends the Pontecorvo formalism to the non-relativistic domain.
We then show how TEUR can be studied in this generalized framework~\cite{Blasone2019}. For consistency with the original notation~\cite{Blasone:1995zc}, in what follows we label neutrino fields with definite masses by latin indices, while greek indices are reserved for definite flavor fields.

\subsection{Time-energy uncertainty relations for neutrino oscillations: Pontecorvo flavor states}
In Section \ref{Unstable} we have re-formulated the TEUR for unstable particles in a second quantization language. The same procedure can be repeated for the Minkowski form of TEUR for neutrino oscillations.

The fields $\nu_j(x)$, $j=1,2$, describe Dirac neutrinos with definite masses $m_j$:
\be
\nu _{j}(x) =  \sum_r \,  \int \!\! \frac{\dr^3 k}{(2 \pi)^{\frac{3}{2}}} \, \left[ u_{{\bf k},j}^{r}(t) \, \alpha _{{\bf k},j}^{r} +  \ v_{-{\bf k},j}^{r}(t)  \,\beta _{-{\bf k},j}^{r\dagger
}\right]  e^{i{\bf k}\cdot {\bf x}}  \, ,
\label{fieldex}
\ee
with $x \equiv ({\bf x},t)$, $\;u^r_{{\bf k},j}(t) =  e^{- i \om_{\G k,j} t}\, u^r_{{\bf k},j}$,
$\;v^r_{{\bf k},j}(t) =  e^{ i \om_{\G k,j} t}\, v^r_{{\bf k},j}$ and 
 $\om_{\G k,j}=\sqrt{|\G k|^2 + m_j^2}$.  
 The ladder operators 
 $\al^r_{\G k, j}$ and $\beta _{{\bf k},j}^{r}$ annihilate the massive neutrino vacuum $|0 \rangle_{1,2}$, i.e.
\be \label{vacm}
\al^r_{\G k, j}|0 \rangle_{1,2} = 0 = \beta _{{\bf k},j}^{r} |0 \rangle_{1,2} \  .
\ee
The anticommutation relations are
\be \label{CAR} \{\nu^{\al}_{i}(x), \nu^{\bt\dag }_{j}(y)\}_{t_x=t_y} =
\de^{3}({\bf x}-{\bf y})
\de _{\al\bt} \de_{ij}\,, 
\ee
\be  \label{CAR2} \{\al ^r_{{\bf k},i}, \al ^{s\dag }_{{\bf q},j}\} = \de
_{\bf k q}\de _{rs}\de _{ij}\,, \,\, \quad \,\, \{\bt^r_{{\bf k},i},
\bt^{s\dag }_{{\bf q},j}\} =
\de _{\bf k q} \de _{rs}\de _{ij}\,,
\ee
while the orthonormality and completeness relations are fixed as
\bea \label{orth}
u^{r\dag}_{{\bf k},i} u^{s}_{{\bf k},i} =
v^{r\dag}_{{\bf k},i} v^{s}_{{\bf k},i} = \de_{rs}\,, \quad\,\, u^{r\dag}_{{\bf k},i} v^{s}_{-{\bf k},i} = 0\,, \quad\,\, \sum_{r}(u^{r\al*}_{{\bf k},i} u^{r\bt}_{{\bf k},i} +
v^{r\al*}_{-{\bf k},i} v^{r\bt}_{-{\bf k},i}) = \de_{\al\bt}\;.
\eea

Similarly to the case of unstable particles, we define the number operator for flavor (Pontecorvo) neutrinos as
\be
N_{P,\si}(t) \ = \ \sum_{\G k,r} \al^{r\dag}_{P,\G k, \si}(t) \al^r_{P,\G k, \si}(t) \,,
\ee
where the flavor ladder operators $\al^r_{P,\G k, \si}$  are just linear combinations of the mass operators
\be \label{tiop}
\al^r_{P,\G k, \si} \ = \ \sum^2_{j=1} \, U_{\si \, j} \, \al^r_{\G k, j} \, .
\ee
Then, Pontecorvo flavor states can be built as
\be \label{postate}
|\nu^r_{\G k, \si}\ran_{P} \ = \ \al^{r\dag}_{P,\G k, \si}|0\ran_{1,2}\,,
\ee
which coincide with Eq.~\eqref{relapp} of the QM treatment when one assumes equal momenta.

The flavor survival probability can be thus computed as the expectation value of the number operator on a reference flavor state
\begin{eqnarray}\label{pontosc}
\mathcal{P}_{\si\rightarrow \si}(t)\ = \ \lan N_{P,\si}(t) \ran_\sigma  \ = \  1 - \sin^2(2\theta)\sin^2 \lf(\frac{\Om_{\G k}^{_-}}{2}t\ri) \, ,
\end{eqnarray}
where, in this subsection, $\langle \cdots\rangle_\sigma = {}_{P}\lan \nu^r_{\G k,\si}| \cdots |\nu^r_{\G k,\si}\ran_P$. Moreover, we have introduced the shorthand notation $\Om_{\G k}^{_\pm} \equiv \om_{\G k,1} \pm \om_{\G k,2}$.
 
It is clear that the number of flavor neutrinos is a good candidate as clock-observable, i.e. $O(t)={N_{P,\si}(t)}$. Noting that
\bea \non
\si^2_N \ = \  \lan N^2_{P,\si}(t) \ran_\sigma \ - \ \lan N_{P,\si}(t) \ran_\si^2
\ = \ \ \mathcal{P}_{\si\rightarrow \si}(t)\lf(1-\mathcal{P}_{\si\rightarrow \si}(t)\ri) \, ,
\eea
we get
\be \label{neutunqm}
\lf|\frac{\dr \mathcal{P}_{\si\rightarrow \si}(t)}{\dr t}\ri| \,\leq  \,2\Delta E  \,\sqrt{\mathcal{P}_{\si\rightarrow \si}(t)\lf(1-\mathcal{P}_{\si\rightarrow \si}(t)\ri)} \, ,
\ee
which is in fact equivalent to Eq.~\eqref{BilTEURNew}. 
An important remark is that $\si_N$ quantifies the dynamic flavor entanglement of the neutrino state~\eqref{postate} (see~\cite{ill2,ill5}) and it naturally appears in the study of Leggett-Garg inequalities in the context of neutrino oscillations \cite{Naikoo:2019gme,Blasone:2022iwf}.

The inequality \eqref{neutunqm} can be made simpler by following similar steps to those below Eq.~\eqref{31}. In so doing, we arrive to
%
%
%
%
\be \label{etrel}
\Delta E\,  T \geq \mathcal{P}_{\si\rightarrow \rho}(T)  \, ,  \quad \si \neq \rho  \, ,
\ee
where $\mathcal{P}_{\si\rightarrow \rho}(t) \ = \  1-\mathcal{P}_{\si\rightarrow \si}(t) $
is the flavor oscillation probability at time $t$. For $T=T_h$ such that $\mathcal{P}_{\si\rightarrow \rho}(T_h)=\ha$, we finally get the analogue of Eq.~\eqref{teurHeis} for neutrino oscillations. 
%
%

\subsection{Flavor Fock space approach to neutrino mixing and oscillations}
Consider the weak decay of a $W^+$ boson, $W^+\rightarrow e^+ + \nu_e$. This process can be described by the Lagrangian 
$\label{Lagrangian} \mathcal{L}=\mathcal{L}_\nu+\mathcal{L}_l + \mathcal{L}_{int}$ with
\bea \lab{neutr}
&&\mbox{\hspace{-2mm}}{\cal L}_{\nu}  =  \overline{\nu}  \lf( i \ga_\mu \pa^\mu - M_{\nu} \ri)\nu \, ,  \\[2mm]
&&\mbox{\hspace{-2mm}}{\cal L}_{l}  = \overline{l} \lf( i \ga_\mu \pa^\mu - M_{l} \ri) l  \lab{lept} \, , \\[2mm]
 &&\mbox{\hspace{-2mm}}{\cal L}_{int}  =  \frac{g}{2\sqrt{2}}
\lf [ W_{\mu}^{+}\,
\overline{\nu}\,\gamma^{\mu}\,(1-\gamma^{5})\,l +
h.c. \ri] \, ,
\label{Linteract}
\eea
where $\nu = \lf(\nu_e, \nu_\mu \ri)^{{T}}  , \, l = \lf(e, \mu \ri)^T$, and
\bea \label{neutmass}
&&\mbox{\hspace{-5mm}} M_{\nu}\,=\,  \lf(\ba{cc}m_e & m_{e \mu}
\\ m_{e \mu} & m_\mu\ea\ri) \, , \qquad
M_l\,=\,  \lf(\ba{cc}\tilde{m}_e &0 \\
0 & \tilde{m}_\mu\ea\ri) .
\eea
The off-diagonal components of $M_{\nu}$ give rise to the neutrino mixing terms in ${\cal L}_{\nu}$.
These terms can be diagonalized by means of the field mixing transformation \cite{Gribov:1968kq,Bilenky:1975tb,Bilenky:1976yj,Bilenky:1977ne,Bilenky:1987ty}
\be  \label{PontecorvoMix1}
\nu_\si(x) \ = \ \sum_{j} \, U_{\si \, j} \nu_j(x) \, ,    ~~\qquad  \sigma = e, \mu;  ~~j = 1, 2\,,
\ee
where $U$ is the mixing matrix
\be
U \ = \ \begin{pmatrix} \cos \theta & \sin \theta \\ -\sin \theta & \cos \theta \end{pmatrix} \, ,
\ee
with $\tan 2 \theta=2 m_{e \mu}/(m_\mu-m_e)$, so that
\bea
{\cal L} & =  & \sum_j \, \overline{\nu}_j  \lf( i \ga_\mu \pa^\mu - m_j \ri)\nu_j  \, + \, \sum_\si \, \overline{l} \lf( i \ga_\mu \pa^\mu - \tilde{m}_{\si} \ri) l  \non \\[2mm]
& + &  \frac{g}{2\sqrt{2}} \sum_{\si,j} \,\lf[ W_{\mu}^{+}(x)\, \overline{\nu}_{j} \, U^{*}_{j \si}\,\gamma^{\mu}\,(1-\gamma^{5})\, l_\si +
h.c. \ri] \, .
\eea
This fact led to develop the idea that neutrino flavor states can be defined as linear combinations of neutrino mass states, as discussed in the previous sections. However, the field mixing transformation \eqref{PontecorvoMix1} does not imply the definition \eqref{postate}. Then, the question arises as to how to consistently define the flavor states $|\nu_\si\ran$ in QFT. 

In order to unveil such point, we start from the observation that the Lagrangian $\mathcal{L}$ is invariant under the action of the global $U(1)$ transformations
$\nu \rightarrow e^{i \alpha} \nu$ and $l \rightarrow e^{i \alpha} l$, 
leading to the conservation of the total  flavor charge $Q_{l}^{tot}$ corresponding to the lepton-number conservation~\cite{Bilenky:1987ty}. This can be written in terms of the flavor charges for neutrinos and charged leptons
\be
Q_{l}^{tot} =  \sum_{\si=e,\mu} Q_\si^{tot}(t) \,,\quad\,\,   Q_{\si}^{tot} (t) = Q_{\nu_{\si}}(t) + Q_{\si}\,,
\ee
with
\bea
Q_{e} & \hspace{-2.2mm}=\hspace{-2.2mm} &  \intx \,
e^{\dag}(x)e(x) \,, \qquad Q_{\nu_{e}} (t) \hspace{-0mm}=\hspace{-0mm}  \intx \,
\nu_{e}^{\dag}(x)\nu_{e}(x)\,,
\nonumber \\ [2mm]
Q_{\mu} &\hspace{-2.2mm} =\hspace{-2.2mm} &   \intx \,
 \mu^{\dag}(x) \mu(x)\,, \qquad Q_{\nu_{\mu}} (t)\hspace{-0mm}=\hspace{-0mm} \intx \, \nu_{\mu}^{\dag}(x) \nu_{\mu}(x)\,  .
 \label{QflavLept}
\eea
Noticing that $[\mathcal{L}_{int}({\bf x},t),Q_\si^{tot}(t)]=0$, we see that neutrinos are produced and detected with a definite flavor~\cite{PhysRevD.45.2414, giunti2007fundamentals}. However, $[(\mathcal{L}_{\nu} + \mathcal{L}_{l}) ({\bf x},t),Q_\si^{tot}(t)] \neq 0$, leading to the flavor oscillation phenomenon.

The second key observation is that the field mixing transformation~\eqref{PontecorvoMix1} can be exactly rewritten as~\cite{ALFINITO199591, Blasone:1995zc}
\bea  
\nu_{e}^{\al}(x)  &=& G^{-1}_{\theta}(t)
\nu_{1}^{\al}( x)
G_{\theta}(t)\,, \\[2mm]
\nu_{\mu}^{\al}(x) &=& G^{-1}_{\theta}(t)
\nu_{2}^{\al}(x) \;
 G_{\theta}(t) \, , 
\eea
where the~\emph{mixing generator} reads
\bea G_\theta(t) & = & \exp[\theta\lf(S_{+}(t) - S_{-}(t)\ri)] \, , \\[2mm]
 S_{+}(t) & \equiv &  \int d^{3}{\bf x}\, \nu_{1}^{\dag}(x) \, \nu_{2}(x) \, , \qquad
S_{-}(t) \ \equiv \ \int d^{3}{\bf x} \,\nu_{2}^{\dag}(x) \, 
\nu_{1}(x) \, .
\eea
In fact, from the above equations we get, e.g., for $\nu_e$
\be
\frac{d^2}{d\theta^2}\,\nu^{\al}_{e}\,=
\,-\nu^{\al}_{e} \, ,
\ee
with the initial conditions
\be
\lf.\nu^{\al}_{e}\ri|_{\theta=0}=\nu^{\al}_{1}\,, \quad\,\,\,\,
\lf.\frac{d}{d\theta}\nu^{\al}_{e}\ri|_{\theta=0}=\nu^{\al}_{2} \, .
\ee

The key point is that the vacuum $|0 \ran_\mass$
 is not invariant
under the action of the mixing generator $G_\theta(t)$. In fact, one has
\be \label{timedep}
|0 (t)\ran_\flav \equiv G^{-1}_\theta(t)\; |0 \ran_\mass
=e^{-\theta\lf(S_{+}(t) - S_{-}(t)\ri)}\, |0 \ran_\mass\,.
\ee
The state~\eqref{timedep} is known as \emph{flavor vacuum} because it is annihilated by the flavor ladder operators $\al_{\sigma}(t)$ and $\bt_{\sigma}(t)$  defined by
\be\al_e(t)|0(t)\ran_\flav \ \equiv \ G^{-1}_\theta(t)\al_1
{ G_\theta(t) \; G^{-1}_\theta(t)}|0\ran_\mass \ = \ 0\,, 
\ee
and similarly for $\bt_{\sigma}(t)$.
Their explicit form is
\bea \lab{operat1}
\al^r_{{\bf k},e}(t)&=&\cos\theta\,\al^r_{{\bf k},1} +
\sin\theta \lf(
 U_{{\bf k}}^{*}(t)\, \al^r_{{\bf k},2}
 + \epsilon^r
V_{{\bf k}}(t)\, \bt^{r\dag}_{-{\bf k},2}\ri)\,, \\
\lab{operat2}
\al^r_{{\bf k},\mu}(t)&=&\cos\theta\,\al^r_{{\bf
k},2}-\sin\theta \lf(
 U_{{\bf k}}(t)\, \al^r_{{\bf k},1}
 - \epsilon^r
V_{{\bf k}}(t)\, \bt^{r\dag}_{-{\bf k},1}\ri)\,,  \\
\lab{operat3}
\!\!\bt^r_{-{\bf k},e}(t)&=&\cos\theta\,\bt^r_{-{\bf
k},1}+\sin\theta\lf(
U_{{\bf k}}^{*}(t)\, \bt^r_{-{\bf k},2}
 -\epsilon^r
V_{{\bf k}}(t)\, \al^{r\dag}_{{\bf k},2}\ri) \,, \\ \lab{operat4}
\!\!\bt^r_{-{\bf k},\mu}(t)&=&\cos\theta \, \bt^r_{-{\bf k},2} -
\sin\theta \lf(
 U_{\bf k}(t)\, \bt^r_{-{\bf k},1}  + \epsilon^r
V_{{\bf k}}(t) \, \al^{r\dag}_{{\bf k},1} \ri) \,. 
\eea
In these equations we have defined $\epsilon^r \equiv (-1)^r$, while $U_{\bf k}$ and $V_{\bf k}$ are the \emph{Bogoliubov
coefficients}
\begin{eqnarray}
U_{{\bf k}}(t)& \equiv & u^{r\dag}_{{\bf k},2}u^r_{{\bf k},1}\;
e^{i(\om_{\G k,2}-\om_{\G k,1})t} \ = \ |U_\G k| \, e^{i(\om_{\G k,2}-\om_{\G k,1})t} \,  \, ,  \\[2mm]
V_{{\bf k}}(t) & \equiv & \epsilon^r\; u^{r\dag}_{{\bf k},1}v^r_{-{\bf k},2}\;
e^{i(\om_{\G k,2}+\om_{\G k,1})t} \ = \ |V_\G k| \, e^{i(\om_{\G k,2}+\om_{\G k,1})t} \, , 
\end{eqnarray}
where the time-independent part of the coefficients is given by
\bea \non
|U_\G k| & \equiv & u^{r\dag}_{{\bf k},2} \, u^{r}_{{\bf k},1} \ = \  v^{r\dag}_{-{\bf k},1} \, v^{r}_{-{\bf k},2} \non \\[2mm]
& = & \left(\frac{\omega_{\G k,1}+m_{1}}{2\omega_{\G k,1}}\right)^{\frac{1}{2}}
\left(\frac{\omega_{\G k,2}+m_{2}}{2\omega_{\G k,2}}\right)^{\frac{1}{2}}
\left(1+\frac{{\bf k}^{2}}{(\omega_{\G k,1}+m_{1})(\omega_{\G k,2}+m_{2})}\right) \, , \label{uk}\\[2mm]
|V_\G k| & = & \epsilon^r\; u^{r\dag}_{{\bf
k},1} \, v^{r}_{-{\bf k},2} \ = \  -\epsilon^r\, u^{r\dag}_{{\bf
k},2} \, v^{r}_{-{\bf k},1} \non \\[2mm]
& = &  \frac{|\G k|}{\sqrt{4 \om_{\G k,1}\om_{\G k,1}}}
\lf(\sqrt{\frac{\om_{\G k,2}+m_2}{\om_{\G k,1}+m_1}}-\sqrt{\frac{\om_{\G k,1}+m_1}{\om_{\G k,2}+m_2}}\ri) \, .
\eea
Notice that $|U_{\bf k}|^2 + |V_{\bf k}|^2 =1$. 
It is straightforward to check that in the relativistic limit $\omega_{\G k,j} \approx |\G k| $, $|U_\G k| \rightarrow 1$ and $|V_{{\bf k}}| \rightarrow 0$. Also, $|V_{{\bf k}}|=0$  when $m_1=m_2$  and/or $\theta=0$, i.e. when no mixing occurs. $|V_{\bf k}|^2$ has the maximum  at $|\G k|=\sqrt{m_1 m_2}$  with $|V_\G k|^2_{max}\ \rar\  1/2$ for $\frac{( m_{2}-m_{1})^2}{m_{1} m_{2}} \rar \infty$, and $|V_{{\bf k}}|^2\simeq \frac{(m_2 -m_1)^2}{4 |\G k|^2}$ for $ |\G k|\gg\sqrt{m_1 m_2}$ at the first non-vanishing order. 

 The flavor fields can be thus expanded as
\begin{eqnarray} \lab{nue}
\nu_{e}(x)
&=& \sum_{{\bf k},r}  \frac{e^{i {\bf k}\cdot{\bf x}}}{\sqrt{V}}
  \lf[ u^r_{{\bf k},1}(t) \,\al^r_{{\bf k},e}(t)\, +\,
v^r_{-{\bf k},1}(t)\,\bt^{r\dag}_{-{\bf k},e}(t)
\ri] \, ,
\\ [2mm]
 \lab{numu}  \nu_{\mu}(x)
&=& \sum_{{\bf k},r}  \frac{e^{i {\bf k}\cdot{\bf x}}}{\sqrt{V}}
\lf[ u^r_{{\bf k},2}(t)\, \al^r_{{\bf k},\mu}(t)\, + \,
v^r_{-{\bf k},2}(t)\,\bt^{r\dag}_{-{\bf k},\mu}(t)
\ri] \, ,
\end{eqnarray}
where $\sum_{{\bf k},r}$ is a simplified notation for the superposition over all field modes of momentum $\textbf{k}$ and polarization $r$. 

A \emph{flavor Hilbert space} (at some reference time, say $t=0$) is defined as 
\be
{\cal H}_{e,\mu} = \lf\{ \al_{e,\mu}^{\dag}\;,\;
\bt_{e,\mu}^{\dag}\;,\; |0\ran_{e,\mu} \ri\},
\ee 
with $|0\ran_{e,\mu} \equiv |0(t=0)\ran_{e,\mu}$.
This is a different Hilbert space with respect to the one of mass-neutrino states. In fact, one can verify that
\be
\lim_{V \rar \infty}\, _\mass\lan0|0(t)\ran_\flav =
\lim_{V \rar \infty}\, e^{\!V \int
\frac{d^{3}{\bf k}}{(2\pi)^{3}}
\,\ln\,\lf(1- \sin^2\theta\,|V_{\bf k}|^2\ri)^2 }= 0\,, \label{ineqrep}
\ee
i.e. flavor and massive fields belong to unitarily inequivalent representations of the anticommutation relations.

The previous discussions suggest that flavor states $|\nu^{r}_{\G k,\si}\ran \ $ can be built as one particle states of the flavor Fock space 
\be \label{bvflavstate}
|\nu^{r}_{\G k,\si}\ran \ = \ \al^{r\dag}_{\G k,\si} |0\ran_{e ,\mu}  \, ,
\ee
and similarly for the antineutrino ($|\bt^{r}_{\G k,\si} \ran \equiv \bt^{r\dag}_{\G k,\si} |0\ran_{e ,\mu}$). One can prove that these states are exact eigenstates of the charge operators at the reference (production/detection) time, i.e.
\be
Q_{\nu_\si}(0) |\nu^r_{\G k,\si}\ran \ = \ |\nu^r_{\G k,\si}\ran \, .
\ee

In this approach the flavor oscillation probability is computed by taking the expectation value of the time-dependent flavor charges with respect to a reference time flavor state~\cite{BHV99}
\be
\mathcal{Q}_{\si\rightarrow \rho}(t) \ = \ \lan Q_{\nu_\rho}(t) \ran_\si \, ,
\ee
where $\langle \cdots\rangle_\si \equiv \lan \nu^r_{\G k,\si}| \cdots |\nu^r_{\G k,\si}\ran$, which gives
\bea \label{oscfor}
&& \mbox{\hspace{-4mm}}\mathcal{Q}_{\si\rightarrow \rho}(t)  =   \sin^2 (2 \theta)\Big[|U_\G k|^2\sin^2\lf(\frac{\Om_{\G k}^{_-}}{2}t\ri)+  |V_\G k|^2\sin^2\lf(\frac{\Om_{\G k}^{_+}}{2}t\ri)\Big]  , \quad \si \neq \rho \, ,  \\[1mm]  \label{oscfor2}
&& \mbox{\hspace{-4mm}}\mathcal{Q}_{\si\rightarrow \si}(t)  =  1 \ - \ \mathcal{Q}_{\si\rightarrow \rho}(t) \, , \quad \si \neq \rho \, .
\eea
Notice the presence of the term proportional to $|V_\G k|^2$ in the oscillation probability Eq.~\eqref{oscfor}, which introduces fast oscillations, that are not present in the usual QM formula (see Eq.~\eqref{pontosc}). As  already mentioned, $|V_\G k|^2 \rar 0$ in the relativistic limit $|{\bf k}| \gg m_j$, $j=1,2$, and the oscillation formula reduces to the standard result, as it should be. It has been proven, in the simple case of scalar field mixing, that the above oscillation formula is the time component of a Lorentz-covariant formula, though the flavor vacuum breaks the Lorentz invariance~\cite{bigs2}. Furthermore, connections between implications of the QFT treatment of mixing and extended (Tsallis-like) statistics have been explored in~\cite{Luciano:2021onl}.

\subsection{Time--energy uncertainty relation for neutrino oscillations in QFT} \label{sectionteurqft}
It is clear that in the QFT treatment lepton charges are natural candidates as clock observables. In fact, starting from
\be
\lf[Q_{\nu_\si}(t) \ , \, H\ri] \ = \ i \, \frac{\dr Q_{\nu_\si}(t)}{\dr t} \ \neq \ 0 \, ,
\ee
we find the {\em flavor--energy} uncertainty relation
\be \label{neutun}
\sigma_H \, \sigma_Q \ \geq \ \frac{1}{2}\lf|\frac{\dr \mathcal{Q}_{\si\rightarrow \si}(t)}{\dr t}\ri|.
\ee
Proceeding as in the Pontecorvo case,  one finds
\be
\lf|\frac{\dr \mathcal{Q}_{\si\rightarrow \si}(t)}{\dr t}\ri| \ \leq \ \Delta E \, .
\ee
From Eq. (\ref{neutun}) we get the simple form for TEUR 
\be \label{etq}
\Delta E \,T \ \geq\  \mathcal{Q}_{\si\rightarrow \rho}(T)  \, ,  \quad \si \neq \rho .
\ee
When $m_i/|\G k|\rightarrow 0$, i.e. in the relativistic case, we get
\bea \label{firstapprox1}
|U_\G k|^2  \approx  1 \  -  \ \varepsilon(\G k)  \, , \;\;\;\;\;
|V_\G k|^2  \approx  \varepsilon(\G k)  \, ,
\eea
with $\varepsilon(\G k)   \equiv {(m_1-m_2)^2}/{4 |\G k|^2}$.
In the same limit
\be
\Om_{\G k}^{_-} \ \approx \ \frac{\delta m^2}{4 |\G k|}\ = \ \frac{\pi}{L^{osc}} \, , \qquad \Om_{\G k}^{_+} \ \approx \ |\G k| \, .
\ee
Evaluating the inequality \eqref{etq} at the leading order and for $T \approx L=L^{osc}/2$, one finds
\be \label{condne1}
\De E \ \geq \ \frac{2 \sin^2(2\theta)}{L^{osc}} \, .
\ee
Then, in such limit we recover the result \eqref{minkteur}. 
As we said above, such neutrino TEUR is usually regarded as a condition for flavor oscillations to occur~\cite{PhysRevD.24.110,Bilenky:2005hv}. In other words, if one managed to measure with a great accuracy the neutrino energy/masses, it could be inferred which massive neutrino was produced in the weak interaction. Then, the oscillation would not occur, {as remarked by Bilenky \emph{et al.}~\cite{Bilenky2008} (see also the discussion on Mossbauer neutrinos in Sec.~\ref{bilteur}).} 
This reasoning is based on the idea that flavor neutrinos are just a superposition of the `'physical" massive neutrinos.
However, the QFT approach bring us to a different interpretation of the inequality (\ref{condne1}).
In fact, note that Eq.~(\ref{ineqrep}) implies
\be \label{neutort}
\lim_{V \rightarrow \infty}\lan \nu^r_{\G k,i}| \nu^r_{\G k,\si}\ran \ = \ 0 \, , \qquad  i=1,2 \, ,
\ee
i.e. neutrino flavor eigenstates, which are produced in charged current weak decays,
\emph{cannot} be generally written as a linear superposition of single-particle massive neutrino states.
However, for Pontecorvo states (\ref{postate}), which are a good approximation of the exact flavor states in the relativistic regime, it holds
\be \label{neutortpon}
\lim_{V \rightarrow \infty}\lan \nu^r_{\G k,1}| \nu^r_{\G k,e}\ran_P \ = \ \cos \theta \, .
\ee
This apparent contradiction is resolved by observing that
\be
\lim_{m_i/|\G k|\rightarrow 0}\,\,\lim_{V \rightarrow \infty} \ \neq \ \lim_{V \rightarrow \infty}\,\, \lim_{m_i/|\G k|\rightarrow 0} \, ,
\ee
i.e. the relativistic ${m_i/|\G k|\rightarrow 0}$ limit cannot be exchanged with the ``thermodynamical'' limit. The relativistic approximation has to be considered just as a single-particle approximation, which does not take into account the intrinsic multi-particle nature of QFT.  The relation~(\ref{neutort}) should be thus understood as
\be
\label{eq125}
\lan \nu^r_{\G k,i}| \nu^r_{\G k,\si}\ran \! =\! \ {}_{1,2}\lan 0_{\G k}|\al^r_{\G k,1} \al^{r \dag}_{\G k,e}|0_{\G k}\ran_{e,\mu}   \prod_{\G p \neq \G k}\!\! {}_{1,2}\lan 0_{\G p}|0_{\G p}\ran_{e,\mu} \, ,
\ee
where we have used the fact that the Hilbert spaces for both massive and flavor fields have a tensor product structure \cite{berezin1966method}. The first factor on the r.h.s. corresponds to~\eqref{neutortpon}, and it is finite. However, as said above, this corresponds to a selection of a single particle sub-space from the complete Hilbert space. In other words, beyond the QM single particle view, the Pontecorvo definition of neutrino state does not work anymore. Then, the inequality (\ref{condne1}) should be regarded as a fundamental bound on the energy precision which can be reached experimentally: as in the case of unstable particles, flavor neutrinos have an intrinsic energy spread which is related with their `'life-time", i.e. $L^{osc}$. {From this perspective, observations of Mossbauer neutrinos could not occur, since the required energy uncertainty would be smaller than the minimum value allowed by TEUR (see the discussion in Sec.~\ref{bilteur})}.

Let us now consider the exact oscillation formula (\ref{oscfor}) at the first order in $\varepsilon(\G k)$:
\begin{eqnarray}
\mathcal{Q}_{\si\rightarrow \rho}(t) \approx \sin^2 (2 \theta) \Big[\sin^2\lf(\frac{\pi t}{L^{osc}}\ri) \lf(1 - \varepsilon(\G k) \ri)  +  \varepsilon(\G k)  \sin^2\lf(|\G k|t\ri)\Big] \, , \quad \si \neq \rho \, .
\end{eqnarray}
Evaluating the inequality \eqref{etq} at $~T= L^{osc}/2$ we get
\be
\label{eq127}
\De E  \ \geq \ \frac{2 \, \sin^2 2 \theta}{L^{osc}} \, \lf[1  -  \varepsilon(\G k) \,  \cos^2\lf(\frac{|\G k|L^{osc}}{2}\ri)\ri] \, .
\ee
We thus find that the energy bound is lowered with respect to \eqref{condne1}.
If we set $m_1=0.0497 \, {\rm eV}$, $m_2=0.0504 \, {\rm eV}$
for neutrino masses in the inverted hierarchy~\cite{PhysRevD.98.030001}
and $|\G k|= 1 \, {\rm MeV}$, we obtain $\varepsilon(\G k) = 2 \times 10^{-19}$, which reveals that such correction is negligible in the relativistic regime and Eq.~\eqref{eq127} can be naively approximated by the quantum mechanical version of TEUR. 

The situation changes if one looks at the non-relativistic regime, e.g., $|\G k|= \sqrt{m_1 m_2}$. In such case,
\bea
|U_\G k|^2 & = & \ha \ +\ \frac{\xi}{2} \ = \  1-|V_\G k|^2 \, , \\[2mm]
\xi & = & \frac{2\sqrt{m_1 m_2}}{m_1+m_2} \, ,
\eea
and we can rewrite the TEUR~(\ref{etq}) in the form
\be
\De E \, T \geq  \frac{\sin^2 2 \theta}{2} \, \lf[1 - \,  \cos \lf(\tilde{\om}_{1}T\ri)\cos \lf(\tilde{\om}_{2}T\ri)  - \, \xi  \sin \lf(\tilde{\om}_{1}T\ri)\sin \lf(\tilde{\om}_{2}T\ri)\ri] \, ,
\ee
where $\tilde{\om}_j = \sqrt{m_j(m_1+m_2)}$. In order to compare it with the previous cases,
we evaluate such expression at $T=\tilde{L}^{osc}/4$, with $\tilde{L}_{osc}=4\pi\sqrt{m_1 m_2}/\de m^2$, obtaining
\be
\De E   \geq  \frac{2\sin^2 2 \theta}{\tilde{L}^{osc}} \ \lf(1-\chi\ri) \, .
\ee
Here we have introduced the shorthand notation
\be
\chi \ = \ \xi \, \sin \lf(\tilde{\om}_{1}\tilde{L}_{osc}/4\ri)\sin\lf(\tilde{\om}_{2} \tilde{L}_{osc}/4\ri) \ + \ \cos \lf(\tilde{\om}_{1}\tilde{L}_{osc}/4\ri)\cos\lf(\tilde{\om}_{2} \tilde{L}_{osc}/4\ri) \, .
\ee
Using the same values for neutrino masses as above, we estimate $\chi=0.1$, which implies that the original bound on the neutrino energy is now decreased by $10\%$.


\section{Conclusion and Discussion}
\label{Disc}

We have discussed recent advances in the study of TEUR for neutrino flavor oscillations. In particular, we have investigated how the the original inequality Eq.~\eqref{final} derived by Bilenky gets changed in the presence of gravity and in a field theoretical picture, respectively. In the first case and for some specific background metrics, it has been shown that gravitational corrections can be rearranged so as to leave the form of the generalized TEUR unchanged, provided one defines an effective (gravity-dependent) oscillation length (see Sec.~\ref{Gravity}). On the other hand, the QFT formulation of TEUR relies upon the identification of the (non-conserved) flavor charge operator with the clock observable. In the latter context, we have argued that the interpretation of neutrinos as ``unstable'' particles does naturally emerge, the life-time being related to the characteristic oscillation length (see the discussion below Eq.~\eqref{eq125}). In passing, we mention that the analogy of oscillating neutrinos with unstable particles has been pointed out also in the recent \cite{Blasone:2023brf}, where the oscillation formula~\eqref{oscfor} has been independently obtained by considering mixing as an interaction and employing the usual perturbation expansion in the Dirac picture. In that work, the role of TEUR is fundamental because it requires the use of finite-time QFT in order to study flavor oscillations, which would be otherwise spoiled in the $S$-matrix formalism ($t \to \infty$).

Further aspects are yet to be explored. In the recent analysis of~\cite{extended} neutrino oscillations have been addressed in extended theories of gravity as a testing ground for the violation of the strong equivalence principle (SEP). Potential violation effects manifest themselves in the form of a generalized oscillation length depending on the SEP parameter. By plugging into the gravity-modified TEUR, one could exploit this result to delve deeper into the interplay between neutrino oscillations and SEP, and possibly constrain SEP corrections. Moreover, a link between cosmic scale phenomena and the QFT properties of field mixing has been analyzed in~\cite{Capolupo:2023fao} in connection with the suggestive interpretation of the flavor vacuum condensate as a dark energy candidate (see also~\cite{Addazi:2022kjt,CarrilloGonzalez:2020oac,Amiri:2021kpp} for other similar dark energy models). It would be interesting to extend our study to this context in the effort to improve the current bounds on neutrino masses through dark energy constraints. On the other hand, growing interest is being aroused by the study of quantum correlations and quantum coherence in neutrino oscillations~\cite{ill2,BLASONE2013320,Banerjee:2015mha, Bittencourt:2022tcl,Li:2022mus,Wang:2023rbf}. Specifically, in~\cite{Bittencourt:2022tcl} the 
complete complementarity relations have been applied to neutrino flavor oscillations to fully characterize the quantumness of such a phenomenon. In line with previous studies, the result has been found that quantum correlations still survive after the complete spatial separation of the
wave packets composing a flavor state, revealing that the quantum nature of mixed neutrinos goes beyond the pure flavor oscillations. 
It is our intent to export TEUR paradigm to the quantum-information theoretic analysis of oscillations to explore the time non-classicality
of this phenomenon in a relativistic domain. 
Preliminary results along this direction appear in~\cite{Blasone:2022iwf}. 
These and some other research lines are under active investigation. 

\acknowledgements
GGL acknowledges the Spanish ``Ministerio de Universidades''
for the awarded Maria Zambrano fellowship and funding received
from the European Union - NextGenerationEU.
He is also grateful for participation to LISA cosmology Working group.
GGL and GL acknowledge networking support from the COST Action CA18108 ``Quantum Gravity Phenomenology in the Multimessenger Approach''. The authors would finally like to thank Al Petrozzi for believing in the original idea and continuous support during the development of the final project.

\bibliography{LibraryNeutrino}

\bibliographystyle{apsrev4-2}

\end{document}